# Erişim İzni Güvenlik Açığı Tespitinde Windows'un Kaynak Tabanlı İzin Mekanizmasının Tersine Bir Yaklaşım

Hakan TEMİZ[1*], Ahmet BÜYÜKEKE[2]

[1]Kontrol ve Otomasyon Teknolojisi, Borçka Acarlar Meslek Yüksekokulu, Artvin Çoruh Üniversitesi, Artvin, 08000
[2]Yönetim Bilişim Sistemleri, İşletme Fakültesi, Adana Alparslan Türkeş Bilim ve Teknoloji Üniversitesi, Adana, 01250

[1]https://orcid.org/0000-0002-1351-7565
[2]https://orcid.org/0000-0002-6103-7646
*Sorumlu yazar: htemiz@artvin.edu.tr

**ÖZ**

Kurum çalışanları görev ve sorumluluklarına göre dosyalarda saklanan bilgilerle çalışırlar. Windows (İşletim sistemi), herhangi bir kullanıcı için herhangi bir iznin kaynak başına ayrı olarak ayarlanması gereken kaynak tabanlı erişim izinlerini kullanır. Bu yöntemde, kaynak ve kullanıcı sayısı arttıkça durum daha karmaşık bir hal alır ve izinlerin atanmasında gözden kaçmalara neden olur. Bu nedenle, herhangi bir çalışanın herhangi bir kaynak kümesi üzerinde hangi izinlere sahip olduğunu incelemek için özel bir mekanizma gereklidir. Bu gereksinim, Windows'un kaynaklara kullanıcı tarafından erişilebilen yaklaşımını tersine çevirerek aşılabilmektedir. Bu yaklaşım, herhangi bir klasörde aktif dizin kullanıcılarına verilen veya reddedilen her türlü iznin hızlı ve kolay bir şekilde incelenmesini ve raporlanmasını sağlayan bir program ile gerçekleştirilmiştir. Araştırmalarımıza göre, yukarıda belirtilen görevleri Windows'un yerleşik araçlarından farklı bir şekilde gerçekleştiren böyle bir araç ilk kez geliştirilmektedir. Önerilen yöntem, yöneticilerin bir güvenlik açığına neden olabilecek eksik veya gözden kaçan bir yetkilendirmenin olmadığından emin olmalarını sağlar. Bu yaklaşım, diğer kaynakları ve diğer yerel veya aktif dizin nesnelerini incelemek için kolaylıkla genişletilebilir.

# An Inverse Approach to Windows' Resource-Based Permission Mechanism for Access Permission Vulnerability Detection

**ABSTRACT**

In organizations, employees work with information stored in files according to their duties and responsibilities. Windows uses resource-based access permissions that any permission for any user has to be set separately per resource. This method gets complicated as the number of resources and users increases, and causes oversights in assigning permissions. Therefore, a special mechanism is required to scrutinize what permissions any employee has on any set of resources. This requirement is circumvented by reversing Windows' approach in terms of user-accessible resources. This approach is implemented by a program allowing quick and easy examination and reporting of any type of permissions granted or denied to active directory users on any number of folders. According to our surveys, this is the first time that such a tool accomplishing above mentioned tasks in a different way than Windows' built-in tools have been developed. The proposed method enables administrators to make sure there is no missing or overlooked setting that could cause a security vulnerability. This approach can easily be extended to scrutinize other resources, and for other local or active directory objects.

## 1. Introduction

It is obvious that, in an organization, users should not have access to any resources not matching their roles and privileges. They may either intentionally or unintentionally collect sensitive information of the company from files and folders, or even going further, corrupt, delete or alter them. Although accidents can and shall happen occasionally, they are often more striking in their effect when caused by administrators. So, administrators have to be more careful to grant or deny permissions to Active Directory (AD) users, groups or services on the resources.

The issues in accessing files or folders are frequently caused by underlying recent modifications, especially denied rights via groups or altered permissions on parent folders (NTFS, 2021). Thus, it is very important to conserve a structured folder hierarchy, especially in a large environment with many users and groups. As the number of users and sources (e.g., files, folders) increases, it gets harder and becomes more complex for administrators to keep access rights under control and to verify everything is okay. Monitoring also becomes more difficult accordingly. After a certain point, oversights and omissions begin in assigning access rights. This leads to unexpected and uncontrollable consequences in accessing resources by users.

The Graphical User Interface (GUI) is Windows' primary mechanism for assigning access permissions, though there are also few command-line tools (e.g., icacls) each serving a specific purpose and not easy to use. These tools are not user friendly as they contain not any GUI. In windows, GUI remains as the first option for administrators to grant, deny and check access permissions on the resources (e.g. folders). However, GUI only allows setting permissions on a particular securable object (SO) and its subfolders via inheritance, if it is a folder. Thus, each SO (e.g., a file, folder, registry key, service or printer) has to be processed individually through GUI.

So, scrutinizing what permissions are granted or denied to any user or group on a list of files or folders turns into a very difficult, cumbersome, and time-consuming job with this GUI. On the other hand, administrators often need to check whether any inappropriate permissions have been granted or denied to a user or group on a set of files or folders. The only way of doing such a task, for example, for multiple folders, is to inspect permission entries via this GUI for each. It is not possible to investigate the access permissions for a single or multiple users or groups on multiple folder or files with a single click. On top of it, the number of SOs can reach tens of thousands. Not any such means exists in GUI to process multiple resources in this manner. Lack of such means makes this job overwhelming, time-consuming and error-prone. Therefore, administrators are forced to fail to properly assign permissions. This article presents a novel approach that addresses the effective access permissions of active directory users in the form of user-based rather than resource-based as Windows addresses. A program is developed to implement this approach. It is tested on a network environment managed by Windows

Server 2012 operating system. It is shown that this approach allows to easily and quickly examine and report any number of permissions allowed or denied to any number of users on any number of folders. With this approach, administrators can seamlessly check if there is any missing or overlooked setting that could cause a security vulnerability by inspecting the active access status of AD users to multiple folders. In this way, administrators of the information technology (IT) department of companies do not waste their valuable time, and can focus on other duties. This approach can be applied for other active directories or local objects with a slight modification to the program.

**2. Related Works**

Administrators of IT departments of companies have many tasks to do in daily routines: network security, license management, antivirus, firewalls, backups, software installation and update, recovery, user support, monitoring, user accounts and permissions, file organization and management, shared folders and so on. Each is very important. Companies use Microsoft's AD to centrally manage and organize many of the above-mentioned tasks and company resources (Binduf et al., 2018). In a typical network, users work with or generate precious information contained in shared files and folders. Every local or AD user, (groups and services as well) should be able to access resources related to their role with appropriate rights and not access other irrelevant resources. Therefore, they must have the appropriate permissions (grants or denials) for these resources, but not more than necessary. This is really critical for mitigating information leakage, data loss (e.g., accidental deletion) and so on. That's why administrators put a lot of effort into assigning rights to users.

In Windows, the security tab in the properties dialog of SOs is the main checkpoint for setting permissions on the resources. Users or groups are authorized or denied certain access roles on a per-resource basis. This means that, for example, each folder has to be handled individually by administrators through this GUI, unless using command-line tools. On the other hand, it is really significant for administrators to make sure permissions are set appropriately on multiple folders: users can access resources they need to reach with proper rights and cannot access the ones they shouldn't. In fact, the effective access tab of the advanced security settings dialog box, shown in Figure 1, opened by clicking "Advanced" button in the security tab of the properties window is designed to scrutinize a user's (or group's) access permissions on a single SO. However, this tool only allows processing a single object at a time. The entire process has to be done iteratively, going through all these dialog boxes for each SO. Conversely, there is no simple and easy way to inspect what kind of access permissions certain users possess on multiple folders, due to the non-existence of such means.

To our best knowledge, there is no such tool that makes up for the deficiency of such a mechanism. Rather, some few command-line tools exist for manipulating access permissions. The first tool that comes to mind is Windows' command line tool, icacls, the successor to the WinNT tool cacls. It runs in command prompt, and is used for displaying or modifying access permissions of files or folders. However, working with it is not so easy since it doesn't have a GUI and requires to learn many of its parameters. In addition, it's not possible to simply get the results in an easily reviewable format, and implement inspections for multiple users and folders. Though it cannot meet the objectives outlined in this article, we demonstrated its output format and deficiencies in the experiments detailed in section 5.

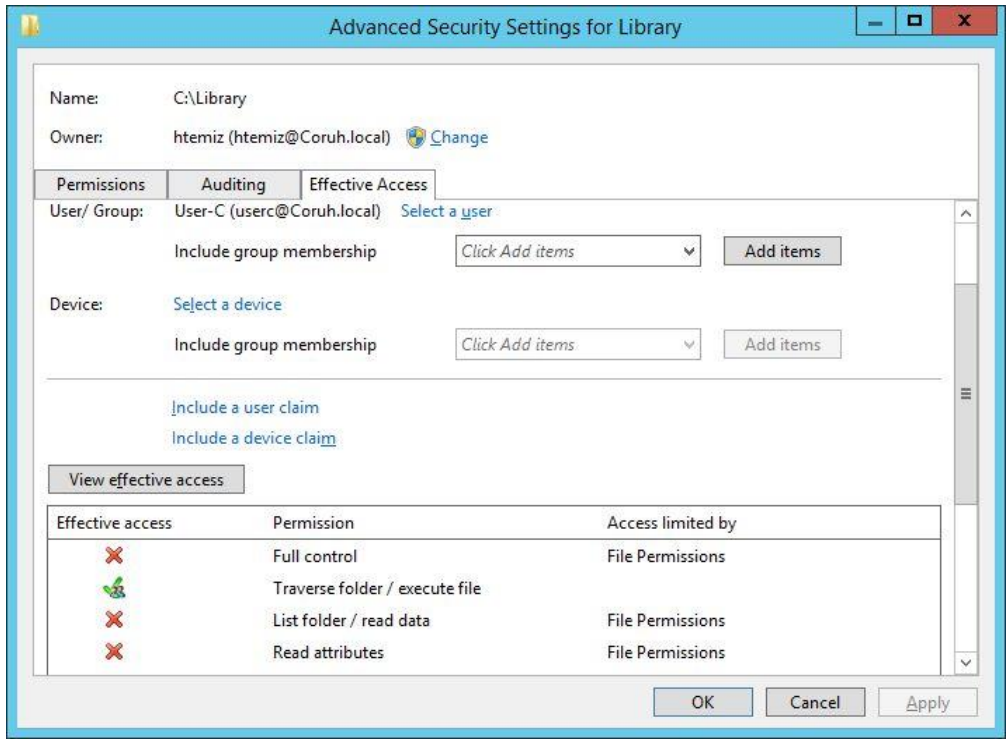

**Figure 1.** Advanced security settings window to examine the effective access permissions assigned to a user or group on a resource.

Sung and Yoon developed a command-line tool in Windows XP environment to secure a single file at a time by assigning desired permissions on it (Sung and Yoon, 2012). Therefore, it does not allow to inspect or check permissions on files or folders, and lacks a GUI. In addition it works only desktop operating system, thus, can only set permissions for local users – Active Directory (AD) users cannot be handled. Another command-line tool is the ACACLS (Cone, 2003). Besides it mimics the operation of WinNT tool cacls, ACACLS has several additional features to extend cacls' functionality. It aimed to edit existing security lists of files and folders, and to manipulate the inheritance flags of files and directories directly. The limitations of Windows's API hindered to fully achieving the second goal. Nevertheless, it is not designed for inspection or reporting the effective permissions. Rather, it allows to set permission with its scripting interface. In addition, it is developed to run only for the users on a local computer, and therefore, it cannot run in a domain environment. Conversely, there is no such tool devised to check effective access permission (EAPs) on SOs of Windows. This gap is filled with new software that implements the approach introduced in this article. It provides administrators an easy and simple way of checking the EAPs of users on multiple folders, with its GUI.

## 3. Windows File and Folder Security

The following subsections provide detailed information on the concepts and components of Windows regarding security including file system, access permission mechanisms.

*3.1. Access Permission Mechanism*

Windows uses two types of security for access control: role-based and Access Control Lists ACL-based. Role-based security is a form of user level security where a server focuses on a logical role of a user rather than focusing on his or her identity. This is simply and commonly implemented with groups that are either local or AD. As opposed to role-based security, ACL-based security focuses more on objects rather than on users. Each SO has its own access control policy represented by a list of permission entries stored in an ACL. It comes with more complexity than role-based system, as typically millions of objects, each with its own ACL, exist in an operating system. The management of this complexity is simplified by inheritance (Brown, 2004).

Access permissions can be assigned one of two ways: explicitly or by inheritance. Explicit permissions are assigned by default when the object is first created, or by user action, whereas inherited ones are given to an object since it is a child of a parent object. Objects inherit all access permissions designated to their containers. If no access is specifically granted or denied, users or groups are denied access (Microsoft, 2009). The precedence hierarchy for the permissions can be summarized as follows -higher precedence permissions are given at the top of the list (Mueller, 2008):

- Explicit Deny
- Explicit Allow
- Inherited Deny
- Inherited Allow

The restrictive permissions override lenient permissions. ACLs allow to set the permissions on files or folders for specific groups or users. ACLs enumerate who (users or groups) has what kind of access (or denials) to certain objects. There are two chief ACL groups: Discretionary Access Control List (DACL) and Security Access Control List (SACL). SACL handles Windows auditing features, whereas The DACL contains a list of access rights granted or denied to certain users or groups on SOs (Microsoft, 2021c).

ACLs convey Access Control Entries (ACEs) that delineate security rights for a user or group. Every typical ACE possesses a header, access mask and security identifier (SID). A SID is a unique value of variable length that is used to identify an individual or group (Microsoft, 2021d). The header defines the type, size, and flags. The access mask specifies the rights that users or groups have to the objects. Inheritance flags in an ACE control how the ACE is to be propagated to the child objects.

Each SO has a security descriptor (SD) that contains all security information related to accessing control for that object. This structure is used to set and query an object's security status. Whenever an object is accessed, its SD is compared to the permissions of accessing user or group to verify that the requested access is allowed. An SD typically includes the following items (Microsoft, 2021b):

- SID of the owner
- SID of primary group
- A DACL
- A SACL
- Qualifiers for the preceding items

The decision to grant or deny access to an SO is made based on the access rights stored in the ACEs in its DACL. When a user's SID matches a specific ACE, ACE is checked to determine if the access type is Allow or Deny. (Microsoft, 2021a) In this process, ACEs are examined in order. As soon as the security system confirms that all requested access entities are granted or that any of them is denied, it yields success in the former case and a failure in the latter.

*3.2. NTFS and Permission Types*

The primary file system of Windows is the NTFS (New Technology File System). There is also a newer Resilient File System (ReFS) suggested for use in Storage Spaces, designed as cost-effective platform maximizing data integrity and availability of very large data sets, as of Windows Server 2012 version. NTFS offers a full set of features covering security descriptors, disk quotas, encryption, and rich metadata, and so on. It can be exploited with Cluster Shared Volumes (CSV) to ensure constantly available volumes which can be accessed concurrently from numerous nodes of a failover cluster (Microsoft, 2017). Windows Server 2012 supports volumes as large as 256 TB depending on the cluster size that can be up to 232-1. NTFS has very robust security features that can be even much higher levels with BitLocker Drive Encryption.

NTFS permissions are set to an SO through ACEs in its ACL (NTFS, 2021). They are logically grouped into six basic permissions, each of which is contained a definite set of advanced (special) permissions. These groups make it easier to set complimentary permissions to users or groups (Stanek, 2008). The entire set of permissions is given in Table 1.

Access rights can be changed in the security tab, as shown in Figure 2, in the properties dialog box opened by selecting the properties option from the pop-up menu appearing when right-clicking on a file or folder.

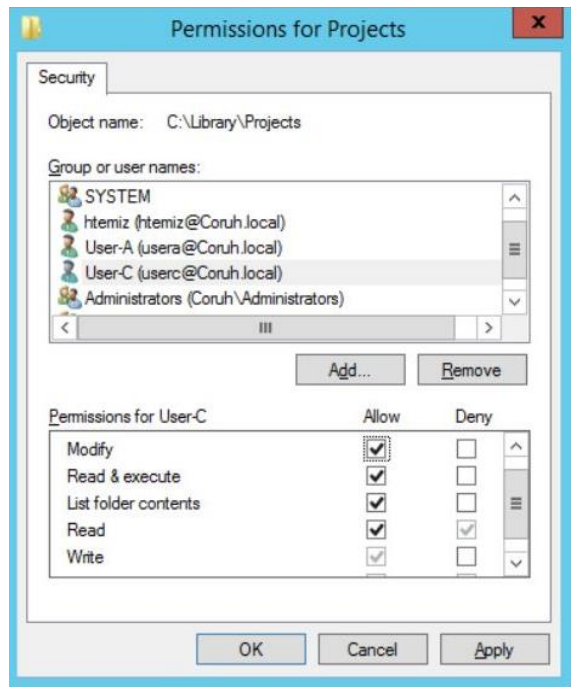

**Figure 2.** Security tab to configure permissions for a file or folder. This screen provides only superficial (not detailed) information of security permissions for a single resource only.

Table 1. Special permissions for folders

| Permission | Full Control | Modify | Read & Execute | List Folder Contents | Read | Write |
|---|---|---|---|---|---|---|
| Traverse Folder / Execute File | ✓ | ✓ | ✓ | ✓ | | |
| List Folder / Read Data | ✓ | ✓ | ✓ | ✓ | ✓ | |
| Read Attributes | ✓ | ✓ | ✓ | ✓ | ✓ | |
| Read Extended Attributes | ✓ | ✓ | ✓ | ✓ | ✓ | |
| Create Files / Write Data | ✓ | ✓ | | | | ✓ |
| Create Folders / Append Data | ✓ | ✓ | | | | ✓ |
| Write Attributes | ✓ | ✓ | | | | ✓ |
| Write Extended Attributes | ✓ | ✓ | | | | ✓ |
| Delete Subfolders and Files | ✓ | | | | | |
| Delete | ✓ | ✓ | | | | |
| Read Permissions | ✓ | ✓ | ✓ | ✓ | ✓ | ✓ |
| Change Permissions | ✓ | | | | | |
| Take Ownership | | | | | | |

## 4. Proposed Work

Windows focuses on SOs rather than users or groups due to the adoption of ACL-based roles in security permissions. Therefore, access permissions are set individually for each user (except inheritance) on each SO. That is, each time only a single SO can be processed unless there is a program or script dedicated to accomplishing this task. Unfortunately, Windows lacks such a tool that simplifying or facilitating the inspection of EAPs on multiple SOs.

Such a task can be accomplished only by developing a special program carefully designed to scrutinize the DACLs of SOs (folders in our case) and report the results in a GUI. Even if it is necessary, results can be recorded in files in any desired format. However, the main purpose of this study is to enable administrators to inspect EAPs in an easy, fast and simple way. This goal is achieved by listing all results in a neat GUI that showing each access type and status (grant or deny) assigned to objects for each user. It is clear that the further functionalities can be included to the program as needed or desired.

Our approach is to recursively traverse the folders and check their ACLs for certain users and permissions being scrutinized. The user can select a single or multiple users from AD, and specify any combination of permission types. The program allows to choose a single folder or multiple folders located within a root folder. It is useful to be able to select multiple folders in a particular location in this way because the folders accessed by network users are usually gathered in a root folder. Certain users are then assigned certain access permissions on specific folders based on their role in the organization. The pseudocode of the algorithm is given below:

**Algorithm** to scrutinize effective access permissions

---
**Inputs**:
    Users: list of active directory users selected from the program
    Folders: path of folders in the root folder selected from the program
    Permissions: list of access permissions to be inspected for each user on each folder
**Output**:
    Table: a data table in tabular format with each row stores True or False for each permission for a user on a folder

**For** each user in Users
        **For** each folder   in Folders
                Insert a row into Table with related columns // (user, folder, and permissions)
    **For** each permission in Permissions
        **For** each ACE in folder's ACL
                **If** ACE belongs to user **OR** a group of which user is member
                      **If** ACE contains the permission **AND** its type is "Allow"
                          Set True in the permission column
                    **Else**
                        Set False in the permission column
        **Else**
            Set False in the permission column
**Return** Table

---

Entire process is very straight. But there are some difficulties in the implementation of this approach because the some feature lack in Windows API. Namely, an ACEs in ACL of a folder may belong either a user or group. If the SID of the ACE is identical to the SID of the user whose permissions are being inspected, we know that that ACE data belongs to that user. We then only check the ACE for the permissions we are inspecting. Otherwise, we need to find out whether the ACE belongs to a group via its SID, and if so, make sure the user is a member of that group to continue further inspections for permissions. Nevertheless, Windows API does not provide a simple mechanism to designate if a SID belongs to a group, or the user is in a role of a group. This issue was overcome with the development of a sub-module that detects membership status by repeatedly checking the members of AD groups.

*4.1. Implementation*

Thanks to Microsoft's .NET framework, many things get much easier than programming with native Windows' API written in low level C++ language. Though it is still required to refer to native libraries for specific tasks from time to time, the .NET elegantly wraps the main important fractions of the functionality of native APIs. As of version 2.0, the .NET framework introduced a new namespace that brings access control programming to the managed C++ API world. Compared to native programming with C++ language, it is simplified very neatly by introducing new methods and classes in the .NET framework.

The program is developed with Microsoft Visual Studio 2015 version using C# language. It is built to run on the .NET framework 4.5 version. However, it can also run on future versions of the .NET framework, provided it is recompiled for the targeted version. For this purpose, it may require a little tweaking to adapt it to the new versions. The program is tested on Windows Server 2012 version with AD on it within a small-scale network which is established with virtualization technology.

The .NET framework provides a wealth of namespaces dedicated to working on certain tasks or technology from desktop to web applications. Each namespace contains a number of classes, methods and other programming elements for a specific scope. This study required to work on AD services, security principles and folder security entries. The details of the main namespaces used for the program are given in Table 2 (Halsey and Bettany, 2015).

**Table 2.** Details of the main namespaces used in the program, which provide basic programming components to perform the tasks related to AD users and access permissions

| Namespace | Remarks |
| --- | --- |
| System.Security.AccessControl | Contains a number of programming elements to manage access control and security-related auditing actions on SOs. The FileSystemSecurity class from this namespace is mainly used to fetch folders' security entries. |
| System.Directory Services. Account Management | Provides a uniform access and management of user, computer, and group security principals through the multiple principal stores such as AD domain services, AD lightweight directory services, and Machine security account manager. |
| System.DirectoryServices | Offers easy access to AD Domain Services from managed code. The namespace covers two different classes: DirectorySearcher and DirectoryEntry. A DirectoryEntry object is constructed to access our Domain and passed to DirectorySearcher class to obtain AD users. |
| System.Security.Principal | Defines the principal object representing the security context under which code is running. Each user are fetched from AD as a Principal object, then, their SIDs are searched in folder's ACEs. |
| System.IO | Allows to read and write to files and data streams, and elementary file and directory support. This namespace is used for typical file and folder operations and to get ACEs of folders with the GetAccessControl method of the Directory() class. |

542

The other base namespaces and classes that are used as core components to develop an application, such as the System.Windows.Forms namespace that used to create the main window of the program, are not mentioned since they are well known. Only the technologies required for the implementation of this particular task are detailed.

The main windows of the program is presented in Figure . The program has two list boxes, one of which lists AD users and the other access types to inspect. Each item in the lists is shown with a checkbox that can be selected individually. Any combination of the items in these lists can be selected. By ticking the checkbox "Select All", all items in a list can be selected or de-selected vice versa. The current folder path is shown next to "Path" text. Users can change the current location by the "Folder Browser" dialog opened when clicking "Change" button. The "Inspect" button kicks inspection process with given settings.

The results are given in a tabular format that each row shows a user's rights to a folder for given access permission types. Each column of access types displays Yes or No to indicate that the user is allowed or denied for the relevant access, respectively. The results can be sorted according to any column in ascending or descending order by just clicking the header of the relevant column. In this way, an administrator can easily examine if the permissions are set properly or not. He or she then be sure that everything is okay, or, takes necessary actions to remedy issues specific to access permissions in case of any overlooked or inappropriate security assignment to users.

Entire results can be stored in an Excel file by clicking the 'Save' button. With this feature, administrators can store access information for later use, further analyze the data using Excel's advanced filtering features, or keep them as a snapshot of settings at a certain time.

We have confirmed through our experiments that our tool correctly identifies the access permissions. For this purpose, we compared the results obtained with the Windows GUI tool (effective access tab) with the results obtained from our tool. We observed that both tools produced exactly the same results.

## 5. Experiment

We established an experiment to demonstrate the impact of our approach. For the experiment we created 10 individual folders under the path C:\Library by assigning each of them a few different permission settings. We created 6 users in our AD environment: User-A, User-B, User-C, User-D, User-E and User-F.

We compared our approach with the icacls command-line tool and the GUI tool of Windows located in the Effective Access tab in the Advanced Security Settings dialog. We measured the completion time only for the Effective Access dialog and our program since it is not possible to accomplish the similar task with the icacls for a reasonable time. We benchmarked these three tools in various aspects in the last part of this section. The following sub-sections present the experiments accomplished with each tool.

*5.1. Experiment with icacls*

In this section, we show how and to what extent an administrator can exploit the icacls.exe tool to check access permissions. As discussed earlier, it is a command-line tool that lists results in some sort of structured text with indentations and uses a special abbreviation for each permission type. Below, we pasted in a snippet of the results when we issued "C:\icacls c:\library\* /t" for the library folder located at the root of C: drive. The t parameter is issued to perform the query for subdirectories.

…

c:\library\Accounts Coruh\Sample Group:(I)(OI)(CI)(DENY)(R)
       Coruh\userc:(I)(OI)(CI)(DENY)(R)
       Everyone:(I)(OI)(CI)(RX)
       Coruh\guess:(I)(OI)(CI)(F)
       Coruh\Sample Group:(I)(OI)(CI)(W)
       Coruh\usera:(I)(OI)(CI)(F)
c:\library\Archive Coruh\Sample Group:(I)(OI)(CI)(DENY)(R)
       Coruh\userc:(I)(OI)(CI)(DENY)(R)
       Everyone:(I)(OI)(CI)(RX)
       Coruh\guess:(I)(OI)(CI)(F)
       Coruh\Sample Group:(I)(OI)(CI)(W)
       Coruh\userb:(I)(OI)(CI)(F)

…

The icacls tool displays or modifies the access permissions with DACLs assigned directly to users or groups on a resource (SO). Each individual permission entry is given with its abbreviation in parenthesis. E.g., (I) and (F) denote the permissions "inherited" and "full access", respectively. In brief, these abbreviations denote the permission types given in

Table **1**, and information about their inheritance status with the letter I. Since it is more technical, domain-specific, and out of the scope of this paper, we keep this information short and concise, and note that interested readers should refer to (Microsoft, 2021c) for further detailed explanations. Each permission entry explicitly assigned to a particular folder is given in the indented rows below the line where the folder path is written.

It's good practice to group users in the same department and/or in the same role into specific groups, and that's how it's usually done. In this way, the operation and management of IT resources are simplified. Access permissions are assigned to groups rather than users unless required. Especially when it is requested to grant/deny a certain access right to a certain user, the relevant regulation is made only for that user.

Given a particular user's access rights are not explicitly assigned for a resource, the only way to resolve the rights of that user is to examine the rights assigned to the groups that that person is a member of. For example, in the last line, it is seen that some permission (I, OI, CI, and F) are explicitly given to "User-A". But what about other users (even more, groups that are members of other groups)? Their access rights are ambiguous, even though there is some preliminary information known by every administrator, e.g., all users are members of the "everyone" group, since they cannot make sure the membership status of a particular user for groups given in DACLs. When an administrator tries to examine the permissions (either, grant or deny) for a particular user (and even in more detail), let us say the User-B, he or she has to check the membership status of User-B recursively for each group given in DACL entry to resolve the inherited rights or denies as described in section 3.1.

In summary, it is far beyond being hard to resolve the detailed information with icacls per user. When it comes to performing such a task for multiple users and resources, no way! However, in fairness, we must state that the icacls is a very useful tool for administrators to set permissions with scripts but not for tasks of which aimed at this paper.

*5.2. Experiment with GUI Tool*

Another way to check access rights of users on particular resources is the Windows GUI served in the Effective Access tab in the Advanced Security Settings dialog, as described in detail in the sections 2 and 3.2. This dialog graphically displays a detailed list for the rights and denies for a particular user on the resource that is right-clicked to open this dialog. The screen output of this tool for the inspection performed for User-B on the Projects folder is given in Figure 3. Detailed output of GUI (effective access tab) when inspection is performed for User-B on the Projects folder.
.

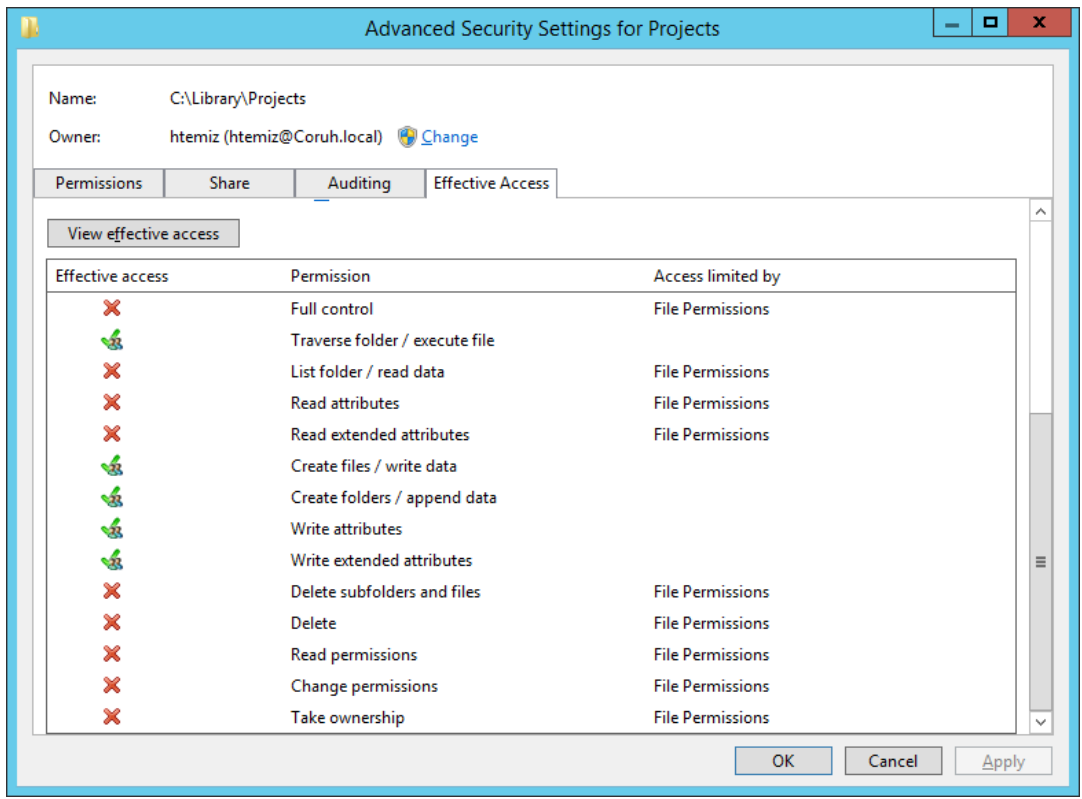

**Figure 3.** Detailed output of GUI (effective access tab) when inspection is performed for User-B on the Projects folder.

To be honest, this dialog box presents the user's access rights in a very neat and detailed way. But only for a single user at a time. One can choose only a single AD or local user each time to display effective rights of that user on the resource. So, it does not provide any means to perform an examination for multiple users and on multiple resources in one go, and export the displayed information in any format. Therefore, performing multiple inspections would turn out to be a very time-consuming, annoying and error-prone task. Given that hundreds of users and many resources exist in organizations, it is not possible to perform such tasks with this tool, as it will take days.

In order to get readers to imagine how long it would take, we asked three administrators to perform our experiment as described at the beginning of this section, and measure the elapsed time. We also asked them to fill the table in the form as yielded by our program given in Table 3. We then averaged their completion times. The average completion time for a single user and 10 folders was 14.47 minutes. If the number of folders remains the same and there were a hundred users, it would take 1447 minutes, in other words, more than 3 business days. However, the number of folders and even users are often many times higher than the numbers assumed in our experience.

The burden this task imposes on the IT department can easily be imagined given the dynamic work environment of IT due to changes in employee roles, entitlements, and even leave and hiring.

*5.3. Experiment with Proposed Tool*

We repeated the same experiment with our tool as well. But this time for more users. In this case, each folder located under the path "C:\library" are inspected for AD users User-A, User-B, User-C and User-D for all permission types. The output of our tool is demonstrated in Figure 4.

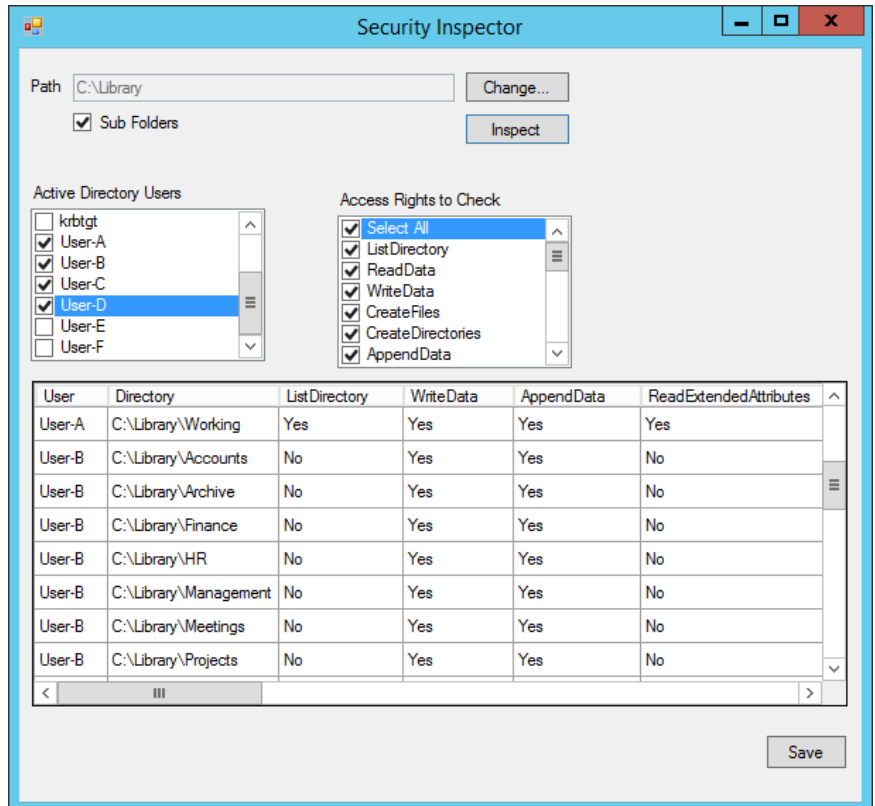

**Figure 4.** Main window of the program and its detailed output when inspection is performed for users User-A, User-B, User-C and User-D on the sub folders of the path C:\Library.

Our tool produces the outputs in a tabular format, which is not possible to obtain with prior discussed Windows' tools. It is very easy with proposed to produce such a detailed report with just few clicks without manual typing, repeating the procedure for multiple users or resources, and so on. Besides diagnosing the results on the display, it is possible to export all reported data to an Excel file. We present in Table 3 a fragment of the report saved in an Excel file for User-B obtained in this experiment. In total, 19 distinct entry info gathered by the tool for each folder and user are presented in a separate columns. If a user has an access right for given permission type, it is denoted as 'Yes' in the relevant column, otherwise as 'No'.

The completion time of our tool is averaged since in the previous experiment the completion time was calculated for one user and for this time four users. The average completion time for 1 user and 10 sub folders is approximately 13 seconds. This is very impressive compared to the completion times of Windows' tools, and considering that there is no need for any manual or repetitive operation to complete the task.

**Table 3.** A fragment of the detailed output of our program written in an Excel file.

| User | Directory | ListDirectory | WriteData | AppendData | ReadExtendedAttributes | WriteExtendedAttributes | Traverse | DeleteSubdirectoriesAndFiles | ReadAttributes | WriteAttributes | Write | Delete | ReadPermissions | Read | ReadAndExecute | Modify | ChangePermissions | TakeOwnership | Synchronize | FullControl |
|---|---|---|---|---|---|---|---|---|---|---|---|---|---|---|---|---|---|---|---|---|
| … | | | | | | | | | | | | | | | | | | | | |
| User-B | C:\Library\Accounts | No | Yes | Yes | No | Yes | Yes | No | No | Yes | Yes | No | No | No | Yes | No | No | No | Yes | No |
| User-B | C:\Library\Archive | No | Yes | Yes | No | Yes | Yes | No | No | Yes | Yes | No | No | No | Yes | No | No | No | Yes | No |
| User-B | C:\Library\Finance | No | Yes | Yes | No | Yes | Yes | No | No | Yes | Yes | No | No | No | Yes | No | No | No | Yes | No |
| User-B | C:\Library\HR | No | Yes | Yes | No | Yes | Yes | No | No | Yes | Yes | No | No | No | Yes | No | No | No | Yes | No |
| User-B | C:\Library\Management | No | Yes | Yes | No | Yes | Yes | No | No | Yes | Yes | No | No | No | Yes | No | No | No | Yes | No |
| User-B | C:\Library\Meetings | No | Yes | Yes | No | Yes | Yes | No | No | Yes | Yes | No | No | No | Yes | No | No | No | Yes | No |
| User-B | C:\Library\Projects | No | Yes | Yes | No | Yes | Yes | No | No | Yes | Yes | No | No | No | Yes | No | No | No | Yes | No |
| User-B | C:\Library\R&D | No | Yes | Yes | No | Yes | Yes | No | No | Yes | Yes | No | No | No | Yes | No | No | No | Yes | No |
| User-B | C:\Library\Surveys | No | Yes | Yes | No | Yes | Yes | No | No | Yes | Yes | No | No | No | Yes | No | No | No | Yes | No |
| User-B | C:\Library\Working | No | Yes | Yes | No | Yes | Yes | No | No | Yes | Yes | No | No | No | Yes | No | No | No | Yes | No |
| … | | | | | | | | | | | | | | | | | | | | |

As a last, we present in Table 4 the comparisons of the tools in terms of operation type, completion time, usability, elaboration, basis and reporting capabilities. Operation indicates if the entire process is accomplished automatically without any user intervention or manually in a repetitive manner. Elaboration means whether the tools give a detailed result or not.

**Table 4.** Comparison of our approach with Windows' command-line and GUI tools. Completion times are computed for a single user and 10 folders.

| Method | Operation | Completion | Usability | Elaboration | Based on | Reporting |
|---|---|---|---|---|---|---|
| Proposed | Automatic | 13 seconds | Very easy | Yes | User | Excel |
| Windows GUI | Manual | 14.47 minutes | Not easy | Yes | Resource | No |
| icacls | Manual | N/A | Too far from easy | No | Resource | No |

As can be seen from the table, the proposed program is superior to the Windows' built-in tools in every respect and provides all kinds of features that these tools cannot provide on their own. It enables access permission inspection to be done much more successfully than Windows can offer, in a short time and with detailed reporting.

## 6. Discussion and Conclusion

In their busy day-to-day work, administrators of IT departments of organizations have to deal with orchestrating users' access to various network resources. It is a very important requirement for them to ensure that the effective access permissions granted or denied to users are appropriate and that there are no vulnerabilities in network security. Windows offers a resource-based access mechanism that makes it possible to check users who have access to a resource on a resource basis (but individually, not collectively). However, being able to list all resources that a user can or cannot access with certain permission types is something needed by administrators to ensure network security. But Windows does not provide such a mechanism to easily and quickly examine the resources accessed on a user basis, together with their access types (grant or denial).

By considering the resource-based access permissions mechanism of Windows from the opposite perspective as user-based access, the proposed approach allowed administrators to examine whether there are any security vulnerabilities in the access permissions assigned to users on multiple resources. In the approach, access permission information is programmatically collected from the resources is transformed into a form of user-based access permission information. With this approach any overlooked, missing, or lenient settings in the assignment of access permissions can easily be found.

As far as we know, a tool that allows detailed inspection and reporting of access permissions in a different way than the existing tools provided by Windows has been developed for the first time. The program allows inspection access rights of any number of AD users on multiple folders in a fast, simple and detailed fashion. The inspections can be done for any combination of access permissions. The experimental studies have revealed that the program will provide a great convenience to administrators in verifying access permissions.

Although this approach has been implemented only for AD users' access permissions to folders, it can be applied to groups and other SOs (e.g. individual files). If necessary, the capabilities of the program can be improved with a few minor adjustments. For example, changing security settings from the GUI, or storing the inspection results in files and so on. Such features can be included in the program through relevant libraries and modules provided in Microsoft .NET framework. Such features can be easily incorporated into the program through the corresponding libraries and modules.

**Conflict of Interest Statement**

The authors of the article declare that there is no conflict of interest.

**Author Contribution Statements**

The authors declare that they have contributed equally to the article.